\documentclass{technical_report}

\usepackage{graphicx} 
\usepackage{xcolor}
\usepackage{listings}

\usepackage{cite} 
\usepackage{array} 
\usepackage{balance} 
\usepackage{caption}
\usepackage{subfig}
\usepackage{fixltx2e} 
\usepackage{hyperref}
\hypersetup{
  colorlinks=true,
  allcolors=black
}
\usepackage{comment}
\newenvironment{techreport}{}{}
\newenvironment{conference}{}{}
\includecomment{techreport}
\excludecomment{conference}

\newcommand{\secref}[1]{Section~\ref{#1}}
\newcommand{\figref}[1]{Figure~\ref{#1}}
\newcommand{\eif}[1]{\lstinline{#1}}

\font\tt=rm-lmtl10
-lmtlo10
-lmtk10
-lmtko10

\begin{document}

\title{Concurrent object-oriented development with behavioral~design~patterns}

\begin{techreport}
\author{%
  Benjamin Morandi$^1$, Scott West$^1$, Sebastian Nanz$^1$, and Hassan Gomaa$^2$
\and
{\small
  \begin{tabular}{cc}
    $^1$ ETH Zurich, Switzerland & $^2$ George Mason University, USA\\
    \texttt{firstname.lastname@inf.ethz.ch} & \texttt{hgomaa@gmu.edu}\\ 
    \texttt{http://se.inf.ethz.ch/} & \texttt{http://mason.gmu.edu/\~{}hgomaa/}\\
  \end{tabular}
}
}  
\end{techreport}

\begin{conference}
\author{%
  Benjamin Morandi\inst{1} \and %
  Scott West\inst{1} \and %
  Sebastian Nanz\inst{1} \and %
  Hassan Gomaa\inst{2} \\
}

\institute{%
  ETH Zurich, Switzerland\\
  \email{firstname.lastname@inf.ethz.ch} \\ 
  \texttt{http://se.inf.ethz.ch/}
  \and
  George Mason University, USA\\
  \email{hgomaa@gmu.edu} \\ 
  \texttt{http://mason.gmu.edu/\~{}hgomaa/}
}
\end{conference}




\maketitle

\begin{abstract}
  The development of concurrent applications is 
  challenging because of the complexity of concurrent designs and the
  hazards of concurrent programming. Architectural modeling using the
  Unified Modeling Language (UML) can support the development process,
  but the problem of mapping the model to a concurrent implementation
  remains.
  This paper addresses this problem by defining a scheme to map
  concurrent UML designs to a concurrent object-oriented program.
  Using the COMET method for the architectural design of concurrent
  object-oriented systems, each component and connector is annotated
  with a stereotype indicating its behavioral design pattern. For each
  of these patterns, a reference implementation is provided using
  SCOOP, a concurrent object-oriented programming model.
  We evaluate this development process using a case study of an ATM
  system, obtaining a fully functional implementation based on
  the systematic mapping of the individual patterns.
  Given the strong execution guarantees of the SCOOP model, which is
  free of data races by construction, this development method eliminates a source of intricate concurrent programming
  errors.
\end{abstract}


\section{Introduction}
\label{sec:introduction}

Writing concurrent applications is no longer a task for specialist
programmers but has become a rather common development task in the age
of multicore computing. Both the complexity of concurrent software
architectures and the hazards associated with concurrent programming,
such as data races and deadlocks, make this task a difficult one.

For concurrent object-oriented applications, support for the
architectural design of the concurrent software is fortunately
available. Standard notations, such as the Unified Modeling Language
(UML), can provide such support when used with a method for
developing concurrent applications, such as
COMET~\cite{gomaa:2000:designing}. 
The remaining
difficulty is the mapping of the concurrent object-oriented model to
an implementation that avoids common concurrency pitfalls.

This paper describes a development method that starts with a concurrent
UML design annotated with behavioral stereotypes and maps the design
systematically to an implementation of the system that is guaranteed
to be data-race free.
Each architectural component in the UML model is given a behavioral
role, based on the COMET object structuring criteria. For each of
COMET's component and connector types we define a mapping to an
implementation in SCOOP (Simple Concurrent Object-Oriented
Programming)~\cite{meyer:1997:oosc,nienaltowski:2007:SCOOP}, 
a concurrent object-oriented programming model. Choosing this model
over others has the benefit of strong execution guarantees: by
construction, the model is free of data races; furthermore, mechanisms for
avoiding deadlocks have been
defined~\cite{west-nanz-meyer:2010:modular}. The mapping of an entire
architectural model to a SCOOP program is based on the mappings of the
individual design artifacts and their composition.

To evaluate the approach, the development process is applied to a case
study of an ATM system that covers all important connector and
component patterns, obtaining a fully functional implementation
of the system.


The remainder of the paper is structured as
follows. \secref{sec:design-patterns} describes behavioral design patterns of the COMET method in UML. After an overview of the SCOOP
concurrency model in \secref{sec:scoop}, the implementation of the
design patterns is described in
\secref{sec:implementation}. \secref{sec:case-study} presents the case
study. 
\secref{sec:related-research} presents a survey of related
work, and \secref{sec:conclusion} draws conclusions.


\newcommand{\stereotype}[1]{{%
\font\larm = larm1000%
\larm%
\char 190}{#1}{%
\font\larm = larm1000%
\larm%
\char 191}}

\section{Behavioral design patterns}
\label{sec:design-patterns}
The concurrent software architecture of a system is best understood by
considering the dynamic characteristics of the system, which is why
this paper focuses on behavioral design patterns. Behavioral design
patterns used by the COMET
method~\cite{gomaa:2000:designing,gomaa:2011:software} 
address design
issues in concurrent and distributed applications.
There are two main
categories of behavioral design patterns. Structural patterns address
the component structure of the concurrent software
architecture. Communication patterns address the message communication
between the concurrent components.

\subsection{Structural patterns and component design}
\label{sec:component-design}

\begin{figure}[htb]
  \centering
  \includegraphics[width=0.8\columnwidth]{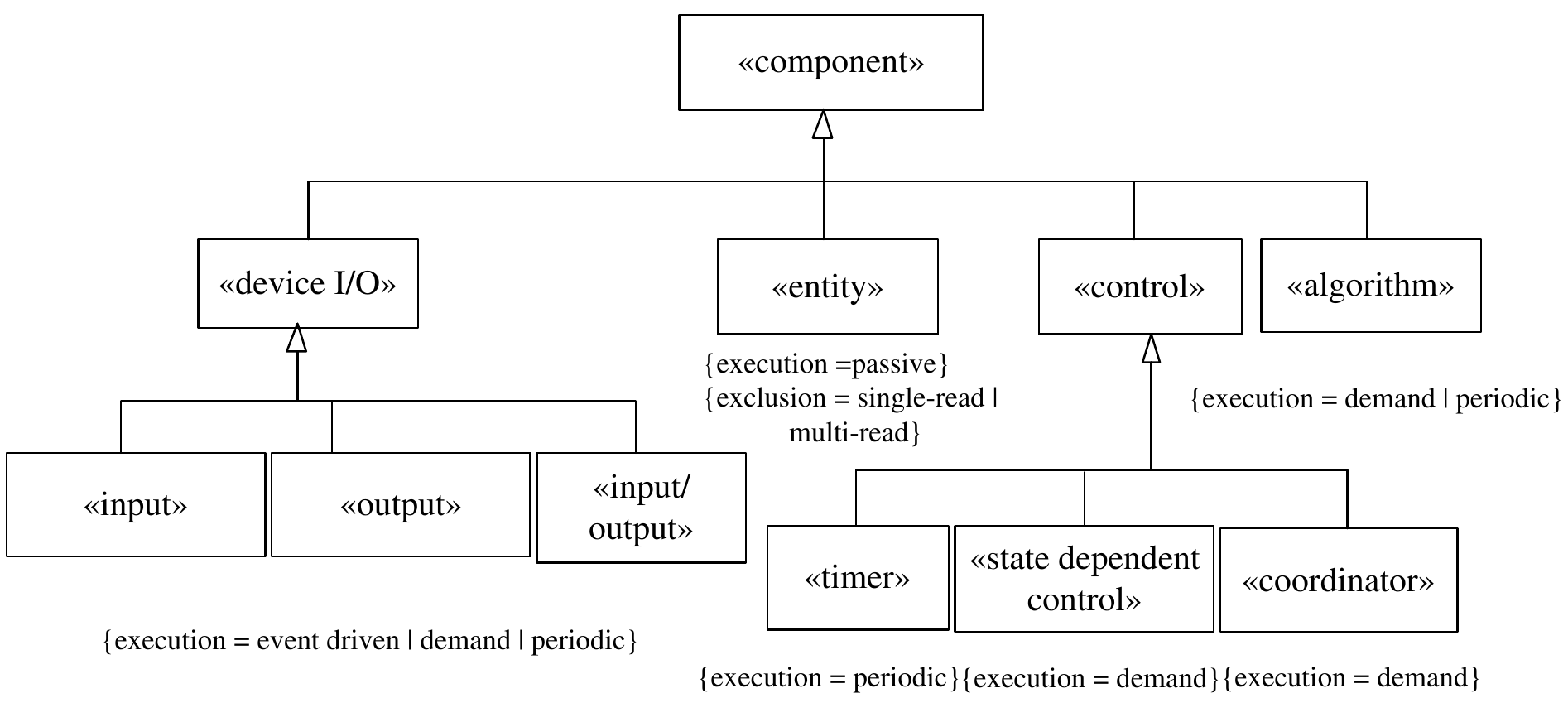}
  \caption{Classification of components using stereotypes}
  \label{fig:components overview}
\end{figure}

Structural patterns address concurrent component design. To assist the
designer, concurrent component structuring criteria are provided. Each
component is depicted from two different perspectives, its role in the
application and the behavioral nature of its concurrency. Models of
the design use UML stereotypes to depict the structuring decisions
made by the designer. The stereotype depicts the component's role
criterion, which describes the component's function in the application
such as \stereotype{I/O} or \stereotype{control}. 
A UML constraint is
used to describe the type of concurrency of the component, which is
based on how the component is activated. For example, a concurrent
\stereotype{I/O} component could be activated by an external event or
a periodic event, whereas an \stereotype{entity} component is passive
and access to it is mutually exclusive or by means of multiple readers
and writers. 
Components are categorized using a component stereotype
classification hierarchy as shown in \figref{fig:components overview}.

\begin{figure}[htb]
  \centering
  \includegraphics[width=0.51\columnwidth]{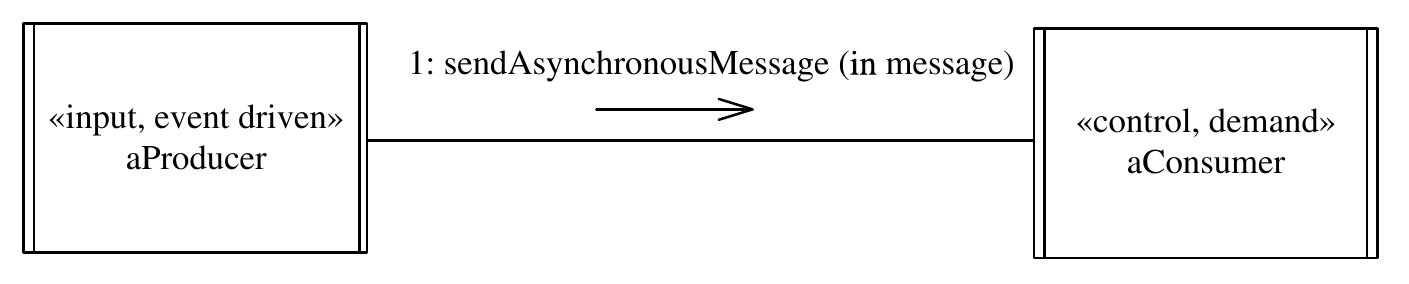}
  \caption{Event-driven input and demand-driven control component}
\label{fig:asynchronous}
\end{figure}

An event driven I/O component interfaces to an event (interrupt)
driven I/O device and is awakened by an external interrupt from the
device. A demand-driven algorithm or control component is
awakened by a message sent by a producer
component. A periodic component is activated by a timer event at regular intervals.
\figref{fig:asynchronous} depicts an event driven input
component communicating with a demand-driven control component. The stereotypes for the producer and consumer depict the component role (e.g., control) followed by the type of concurrency (e.g., demand). 
Alternatively, separate stereotypes could be used to depict the component role and the type of concurrency, which is supported by UML 2~\cite{gomaa:2011:software} 
but not by all UML tools.

\subsection{Communication patterns and connector design}
\label{sec:connector-design}

Communication between concurrent components is a particularly
important design issue because, unlike sequential
systems in which call/return is the only pattern of communication
between sequential components, there are many different ways in which
concurrent components can communicate with each other.  Communication
patterns describe the different types of message
communication between the concurrent components of the software
architecture. In both distributed and non-distributed applications,
communication patterns include asynchronous communication and
synchronous communication with or without reply.
Other communication
patterns are exclusively for distributed systems, including brokered
communication, event-based communication, and transaction-based
communication. 

\begin{figure}[htb]
  \centering
  \includegraphics[width=0.35\columnwidth]{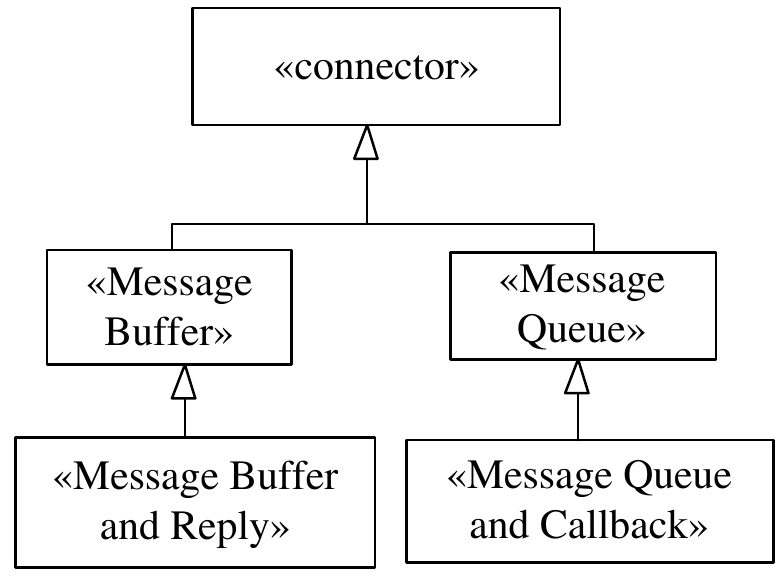}
  \caption{Classification of connectors using stereotypes}
  \label{fig:connectors overview}
\end{figure}

Using the component/connector paradigm, a
connector can be designed for each communication pattern to
encapsulate the details of the communication mechanism. The \emph{message buffer} and \emph{message buffer and reply} connectors implement the synchronous communication pattern without respectively with reply; the \emph{message queue} and \emph{message queue and callback} connectors implement the corresponding asynchronous communication patterns. These connectors can also be categorized using a stereotype classification hierarchy as shown in \figref{fig:connectors overview}.

\begin{figure*}[htb]
\centering
  \begin{tabular}{@{}cc}    
\subfloat[Message buffer connector for synchronous message communication without reply pattern]{
  \centering
  \includegraphics[width=0.44\columnwidth]{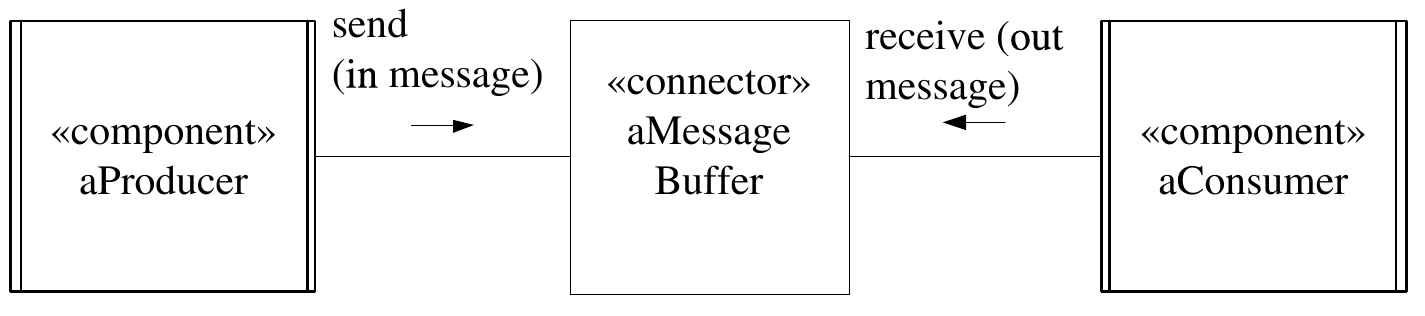}
  \label{fig:connector-synchronous-no-reply}
}
&
\subfloat[Message queue connector for asynchronous message communication pattern]{
  \includegraphics[width=0.5\columnwidth]{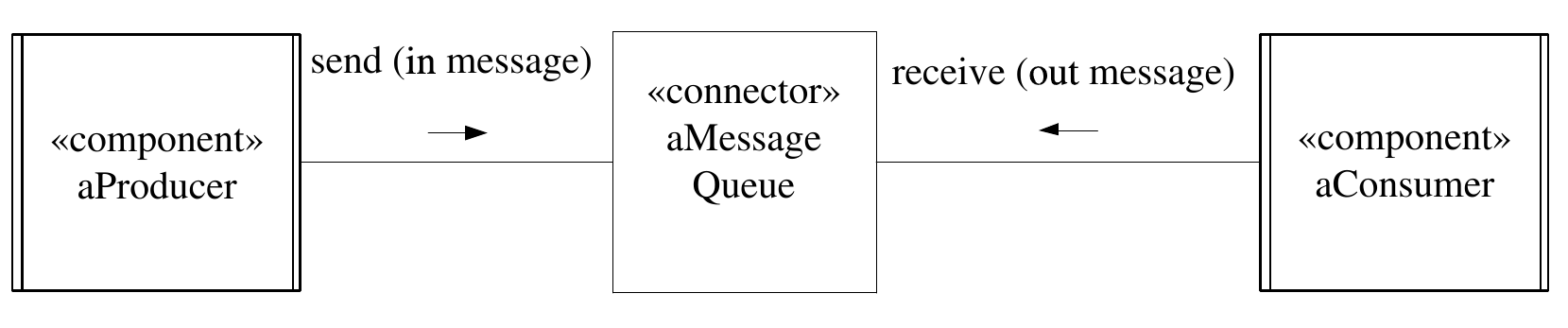}
  \label{fig:connector-asynchronous}
}
\\
\subfloat[Message buffer and reply connector for synchronous communication with reply pattern]{
  \includegraphics[width=0.44\columnwidth]{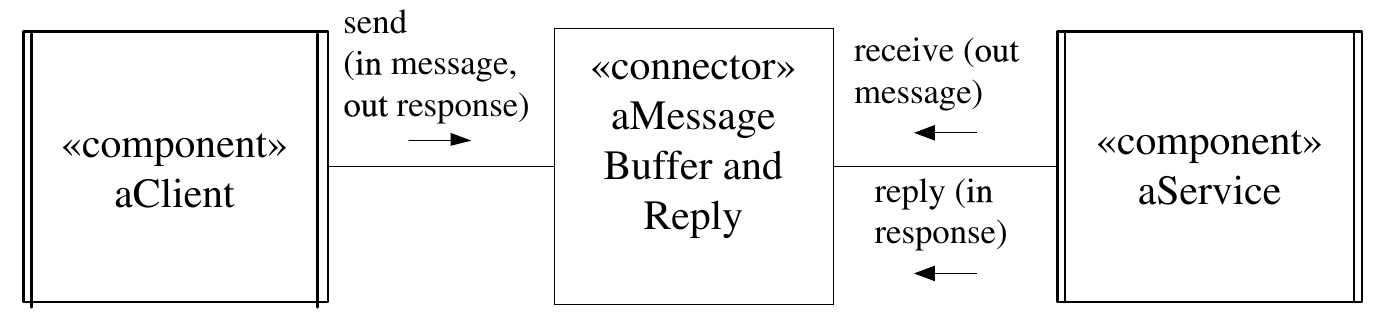}
  \label{fig:connector-synchronous-reply}
}
&
\subfloat[Message queue and callback connector for asynchronous communication with callback pattern]{
  \centering
  \includegraphics[width=0.5\columnwidth]{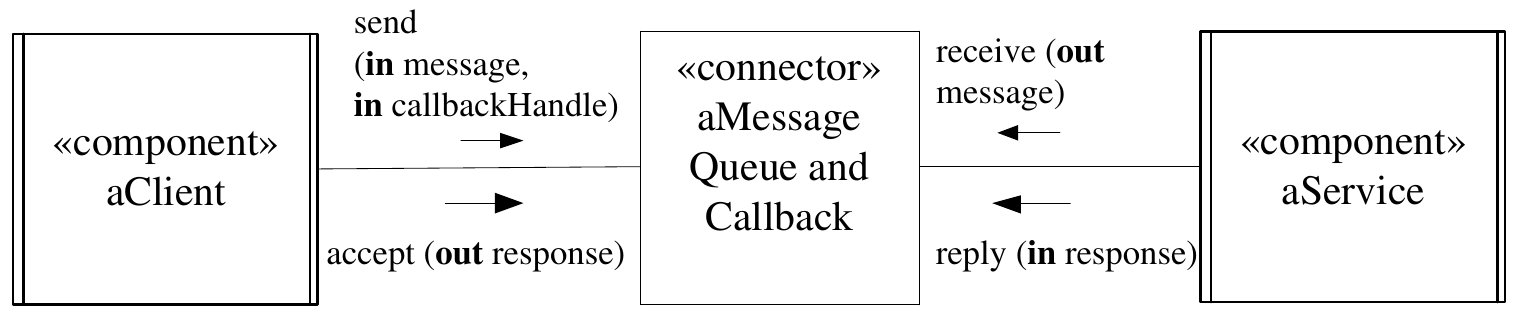}
  \label{fig:asynchronous-callback-handle}
}
  \end{tabular}
  \caption{Connectors for communication patterns}
  \label{fig:connectors for communication patterns}
\end{figure*}

\figref{fig:connector-synchronous-no-reply} depicts a synchronous
communication without reply pattern, in which the concurrent producer component sends a
message to a concurrent consumer component via a message buffer connector, and waits for the consumer to accept the message.
\figref{fig:connector-asynchronous} depicts an asynchronous message
communication pattern in which a producer communicates with a consumer
through a message queue connector that encapsulates the details of the asynchronous
communication by: (1) adding a message from the producer to a FIFO
message queue and only suspending the producer if the queue is full
(2) returning a message to a consumer or suspending the consumer if
the queue is empty. 
\figref{fig:connector-synchronous-reply} depicts a synchronous
communication with reply pattern in which the client component sends a message
to a service component and waits for the reply via a message buffer and reply connector.
\figref{fig:asynchronous-callback-handle} depicts an asynchronous
communication with reply pattern in which the client sends a
message to a service via a message queue and callback connector,
continues executing and later receives the service response from the
connector. In this pattern, the client needs to provide an id or
callback handle at which the response is returned, as shown in
\figref{fig:asynchronous-callback-handle}. 
Note that the callback
response is handled asynchronously, and hence differently from the
synchronous communication with reply pattern. For this reason, the
response is named a callback to distinguish it from a synchronous
reply.

\section{SCOOP}
\label{sec:scoop}
The main idea of SCOOP \cite{meyer:1997:oosc,nienaltowski:2007:SCOOP,morandi:2010:scoop} 
is to simplify the writing of correct concurrent programs, by allowing developers to use familiar concepts from object-oriented programming while protecting them from common concurrency errors such as data races. 
%
Empirical evidence supports the claim that SCOOP indeed simplifies reasoning about concurrent programs as opposed to more established models~\cite{nanz-torshizi-pedroni-meyer:2011:usability_of_SCOOP}. SCOOP has been developed on top of Eiffel, an object-oriented programming language; 
however, SCOOP's concurrency model can be applied to other object-oriented programming languages, for example to Java \cite{torshizi-ostroff-paige-chechik:2009:JSCOOP}.

In SCOOP, every object is associated for its lifetime with a processor, called its \emph{handler}. A \emph{processor} is an autonomous thread of control capable of executing actions on objects. An object's class describes the possible actions as \emph{features}. 
%
A variable \lstinline[language=SCOOP]!x! belonging to a processor can point to an object with the same handler (\emph{non-separate object}), or to an object on another processor (\emph{separate object}). In the first case, a \emph{feature call} \lstinline[language=SCOOP]!x.f! is \emph{non-separate}: the handler of \lstinline[language=SCOOP]!x! executes the feature synchronously. In this context, \lstinline[language=SCOOP]!x! is called the \emph{target} of the feature call. In the second case, the feature call is \emph{separate}: the handler of \lstinline[language=SCOOP]!x!, i.e., the \emph{supplier}, executes the call asynchronously on behalf of the requester, i.e., the \emph{client}. The possibility of asynchronous calls is the main source of concurrent execution. 

The producer-consumer problem serves as a simple illustration of these ideas. A root class defines the entities \lstinline[language=SCOOP]!producer!, \lstinline[language=SCOOP]!consumer!, and \lstinline[language=SCOOP]!buffer!. 
\begin{lstlisting}[mathescape=true,language=SCOOP]
producer: separate PRODUCER
consumer: separate CONSUMER
buffer: separate BUFFER [INTEGER]
\end{lstlisting}
The keyword \lstinline[language=SCOOP]!separate! specifies that the referenced objects may be handled by a processor different from the current one. A \emph{creation instruction} on a separate entity such as \mbox{\lstinline[language=SCOOP]!producer!} will create an object on another processor; by default the instruction also creates that processor.

Both the producer and the consumer access an unbounded buffer in feature calls such as \lstinline[language=SCOOP]!buffer.put (n)! and \mbox{\lstinline[language=SCOOP]!buffer.item!.} To ensure exclusive access, the consumer must lock the buffer before accessing it. Such locking requirements of a feature must be expressed in the formal argument list: any target of separate type within the feature must occur as a formal argument; the arguments' handlers are locked for the duration of the feature execution, thus preventing data races. Such targets are called \emph{controlled}. For instance, in \lstinline[language=SCOOP]!consume!, \lstinline[language=SCOOP]!buffer! is a formal argument; the consumer has exclusive access to the buffer while executing \lstinline[language=SCOOP]!consume!.

\begin{lstlisting}[language=SCOOP]
consume (buffer: separate BUFFER [INTEGER])
		-- Consume an item from the buffer.
	require 
		not (buffer.count = 0)
	local 
		consumed_item: INTEGER
	do 
		consumed_item := buffer.item 
	end
\end{lstlisting}

Preconditions (after the \lstinline[language=SCOOP]!require! keyword) express \emph{wait conditions}; any precondition of the form \lstinline[language=SCOOP]!x.some_condition! makes the execution of the feature wait until the condition is true. For example, the precondition of \lstinline[language=SCOOP]!consume! delays the execution until the buffer is not empty.
As the buffer is unbounded, the corresponding producer feature does not need a precondition.

\vspace{2ex}

\begin{lstlisting}[language=SCOOP]
produce (buffer: separate BUFFER [INTEGER])
		-- Produce an item and put it into the buffer.
	local
		produced_item: INTEGER
	do
		produced_item := create {INTEGER}.make_random
		buffer.put (produced_item)
	end
\end{lstlisting}

The runtime system ensures that the result of the call \mbox{\lstinline[language=SCOOP]!buffer.item!} is properly assigned to the entity \mbox{\lstinline[language=SCOOP]!consumed_item!} using a mechanism called \emph{wait by necessity}: while the consumer usually does not have to wait for an asynchronous call to finish, it will do so if it needs the result.


\section{Implementation of design patterns}
\label{sec:implementation}

This section describes the SCOOP implementation of the behavioral
design patterns with examples, and highlights the most relevant
implementation properties. The full implementation is available
online~\cite{design_patterns_github:2012:ETH}.

\subsection{Implementing components}

Components are implemented by providing a class hierarchy
mirroring the component taxonomy in \figref{fig:components overview}.
Specialized components in the end user application
inherit from the appropriate abstract class.
This approach allows for a wide variety of
component implementations to be accommodated in the same hierarchy.
To remove ambiguity, 
the term \emph{component object} will be used to denote
an instance of the \emph{component class},
which is the implementation of the design pattern component.

We examine the implementation of one component, the periodic task,
in detail.
The periodic task component is implemented as a pair of classes:
one class represents the job to be done,
the other is a ``pacemaker'' which periodically calls the 
first class to perform its task.
The instances of each class should reside
on two distinct processors.

\begin{figure}[htb]
  \begin{lstlisting}[language=SCOOP]
deferred class PERIODIC
feature
  is_done: BOOLEAN
  step deferred end
feature {PACEMAKER}
  notify do ... end
end
  \end{lstlisting}
  \caption{Periodic class interface}
  \label{fig:periodic}
\end{figure}

The basic interface to \eif{PERIODIC} can be seen in 
\figref{fig:periodic}.
The abstraction defines:
\begin{itemize}
\item a single iteration (\eif{step}),
\item an indicator that the task is finished (\eif{is_done}),
\item integration with the pacemaker: 
  \eif{notify} executes a step then asks the pacemaker to
  schedule another call to \eif{notify} (unless \eif{is_done}).
\end{itemize}

This design increases the availability of
the \mbox{\eif{PERIODIC}} object to other processors.
If the waiting (via \mbox{\eif{sleep})} were to occur within the 
\eif{PERIODIC} object,
that object's processor would be unavailable for the duration
of the \eif{sleep} routine;
other objects would be unable to ask the periodic task simple
queries such as \eif{is_done}.
This is why the pacemaker does the waiting and
calls to the task after an appropriate delay.
The interaction between the pacemaker and the
periodic task allows the processor containing the
periodic task to remain unoccupied between \eif{step} executions.

\subsection{Implementing connectors}
Each of the connectors in \figref{fig:connectors overview} is implemented
using three highly dependent pieces:
the sender endpoint, the receiver endpoint, and the conduit(s).
These are implemented as a cohesive unit to 
guarantee the communication takes place correctly.
The sender and receiver endpoints are responsible for
ensuring that the appropriate protocol is obeyed.
Conduits are data channels; they sit as a bridge between endpoints,
with the endpoints responsible for using the conduit correctly
(e.g., ensuring synchronous access).
We use the term \emph{connector objects} to denote the combination
of \emph{endpoint objects} and \emph{conduit objects},
which form the realization of a particular connector.

An example of a simple connector is the synchronous message buffer.
It holds a single message and the sender does not proceed until
the receiver has received the message.
The implementation of the message buffer conduit can be seen
in~\figref{fig:sync_conduit}.

\begin{figure}[htb]
  \begin{lstlisting}[language=SCOOP]
class SB_CONDUIT [G]
feature
  put (msg: G)
    require 
      is_empty
    do 
      create cell.put (msg) 
    end
  remove: G
    require 
      not is_empty 
    do 
      Result := cell.item
      cell := Void 
    end
  is_empty: BOOLEAN
    do 
      Result := (cell = Void) 
    end
feature {NONE}
  cell: CELL [G]
end
  \end{lstlisting}
  \caption{Conduit class for message buffer connector}
  \label{fig:sync_conduit}
\end{figure}

\noindent The conduit contains a storage container with space for a single object of any type, \eif{G}, 
and routines to set, clear, and report on the contents of the container.
The core behavior of the connector is encoded in the endpoints.
In \figref{fig:sync_sender_receiver}, the sender
guarantees it only proceeds after the receiver has removed the message
by using the wait condition \mbox{\eif{conduit.is_empty}}
in the \eif{rendez_vous} routine.
Likewise, the receiver waits on
message arrival.
The implementation details are hidden from the external
interface of the end points, 
only the \eif{send} and \eif{receive} routines
are exposed,
making the interface%
\begin{techreport}
, \figref{fig:scoop_buffer_class_diagram}, 
\end{techreport}
 quite simple. The usage of this connector can be seen in the
object diagram in \figref{fig:scoop_buffer_component_diagram},
which is the SCOOP implementation of
\figref{fig:connector-synchronous-no-reply}.

\begin{figure}[htb]
  \begin{lstlisting}[language=SCOOP]
class SB_SENDER [G] create make
feature
  send (msg: G)
    do
      sep_send (msg, send_conduit)
      rendez_vous (send_conduit)
    end
feature {NONE}
  send_conduit: separate SB_CONDUIT [G]
  make (a_send_conduit: separate SB_CONDUIT [G])
    do
      send_conduit := a_send_conduit
    end
  sep_send (msg: G; conduit: separate SB_CONDUIT [G])
    require 
      conduit.is_empty
    do 
      conduit.send (msg) 
    end
  rendez_vous (conduit: separate SB_CONDUIT [G])
    require 
      conduit.is_empty
    do end
end

class SB_RECEIVER [G] create make
feature
  receive: G
    do 
      Result := sep_receive (recv_conduit) 
    end
feature {NONE}
  recv_conduit: separate SB_CONDUIT [G]
  make (a_recv_conduit: separate SB_CONDUIT [G])
    do
      recv_conduit := a_recv_conduit
    end
  sep_receive (conduit: separate SB_CONDUIT [G]): G
    require 
      not conduit.is_empty
    do 
      Result := conduit.remove 
    end
end
  \end{lstlisting}
  \caption{Sender and receiver endpoint classes for message buffer connector}
  \label{fig:sync_sender_receiver}
\end{figure}

\begin{techreport}
\begin{figure}[htb]
\centering
\subfloat[Object diagram for conduit and endpoints]{
  \includegraphics[width=0.6\linewidth]{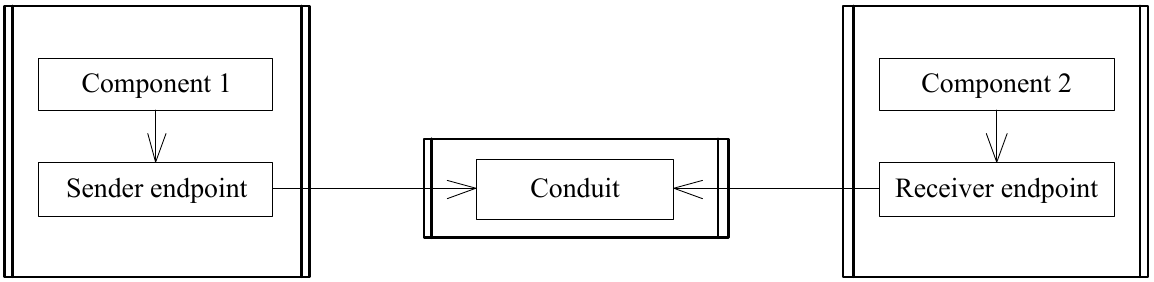}
  \label{fig:scoop_buffer_component_diagram}
}

\subfloat[Class diagram for conduit and endpoints]{
  \includegraphics[width=0.7\linewidth]{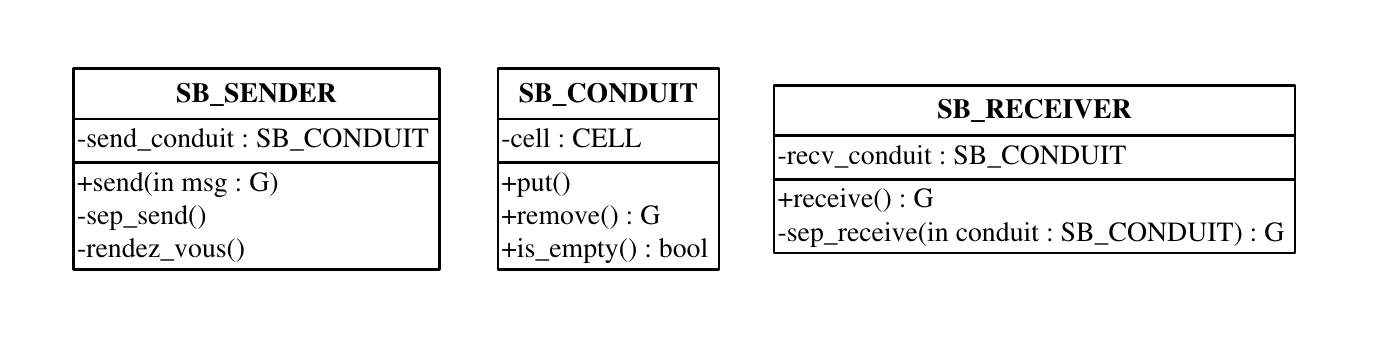}
  \label{fig:scoop_buffer_class_diagram}
}
\caption{Connectors for synchronous buffer}
\end{figure}
\end{techreport}

\begin{techreport}
\begin{figure}[!htb]
\centering
\subfloat[Object diagram for conduit and endpoints]{
  \includegraphics[width=0.6\linewidth]{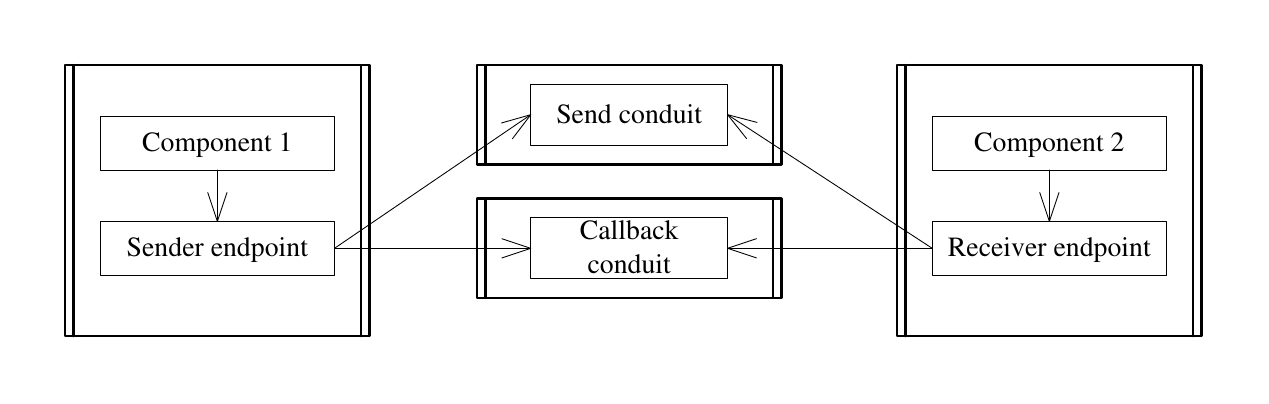}
  \label{fig:scoop_queue_component_diagram}
}

\subfloat[Class diagram for conduit and endpoints]{
  \includegraphics[width=\textwidth]{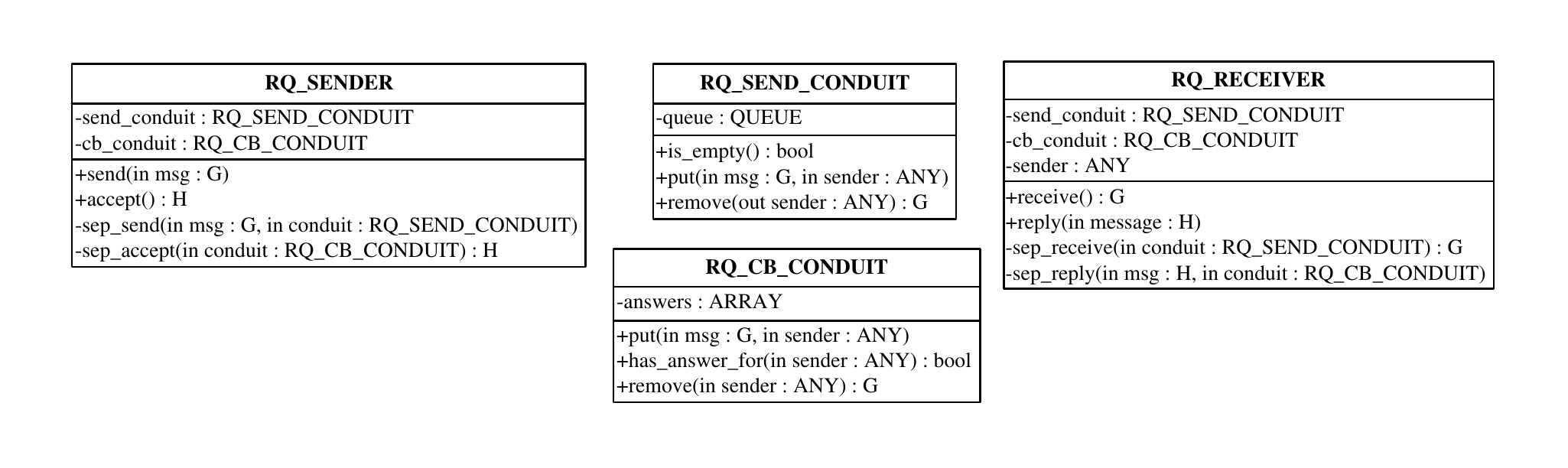}
  \label{fig:scoop_queue_class_diagram}
}
\caption{Connectors for asynchronous queue with callback}
\end{figure}
\end{techreport}

\begin{conference}
\begin{figure}[!htb]
\centering
\begin{tabular}{c}
\subfloat[Message buffer connector]{
  \includegraphics[width=0.6\linewidth]{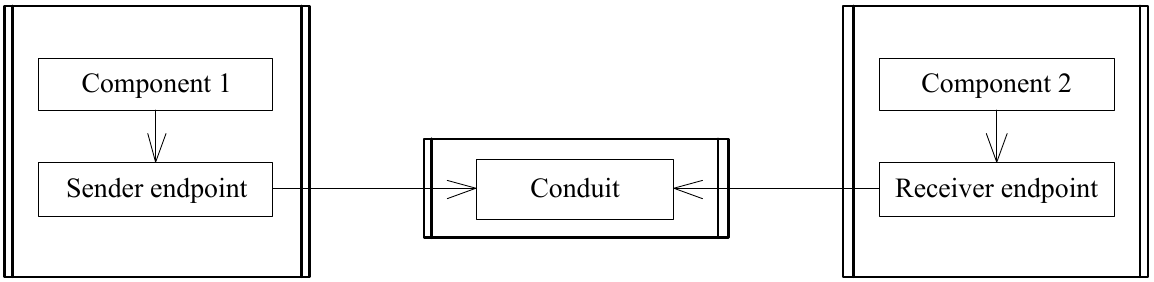}
  \label{fig:scoop_buffer_component_diagram}
}\\
\subfloat[Message queue and callback connector]{
  \includegraphics[width=0.6\linewidth]{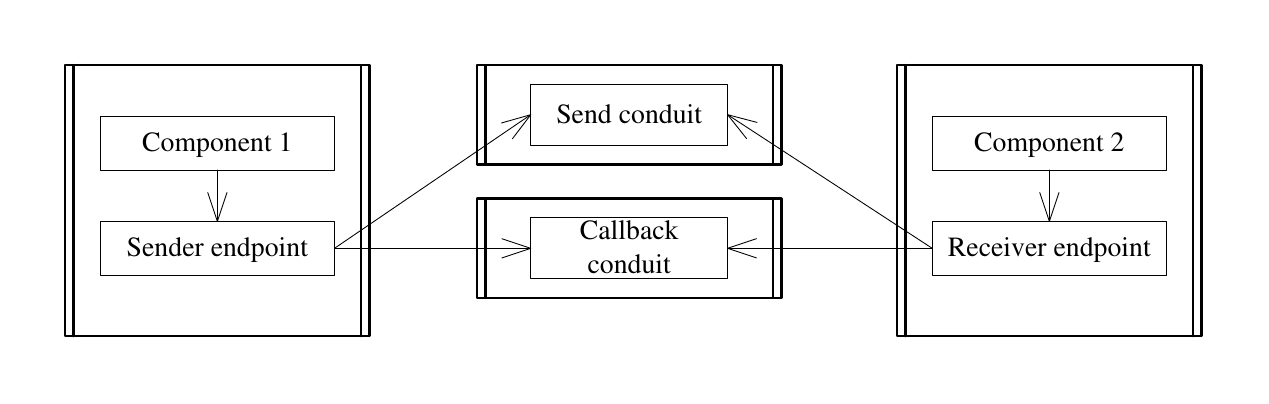}
  \label{fig:scoop_queue_component_diagram}
}  
\end{tabular}
\caption{Object diagrams for conduit and endpoints}
\end{figure}
\end{conference}

Another example is an asynchronous message queue with callback,
where the sender sends its message, continues on, 
then waits for a reply.
The queue with callback connector 
sends and receives messages of type \eif{G} and \mbox{\eif{H},} respectively.
This is seen in the implementation given in \figref{fig:scoop_queue_component_diagram}, which is the SCOOP implementation of
\figref{fig:asynchronous-callback-handle}.

The connector is implemented using two independent conduits;
one conduit is responsible for carrying outgoing messages and
the other for replies
(this pattern is common in connectors with reply).
The sender uses the conduits in two basic ways:

\begin{itemize}
\item Sending a message, 
  along with its identity. 
  This allows the receiving end to send a message back to it.
\item Receiving the callback from the other end.
  The sender's identity is used once again to select the
  correct message to receive.
\end{itemize}

\begin{techreport}
\begin{figure}[htb]
\begin{lstlisting}[language=SCOOP,captionpos=b]
class RQ_SENDER [G, H] create make
feature
  send (msg: G)
    do 
      sep_send (msg, send_conduit) 
    end
  accept: separate H
    do 
      Result := sep_accept (cb_conduit) 
    end
feature {NONE}
  send_conduit: separate RQ_SEND_CONDUIT [G]
  cb_conduit: separate RQ_CB_CONDUIT [H]
  make (a_send_conduit: RQ_SEND_CONDUIT [G];
        a_cb_conduit: RQ_CB_CONDUIT [H])
    do
      send_conduit := a_send_conduit
      cb_conduit := a_cb_conduit
    end
  sep_send 
    (msg: G; conduit: separate RQ_SEND_CONDUIT [G])
    require 
      not conduit.is_full
    do 
      conduit.put (msg, Current) 
    end
  sep_accept: separate H 
    (conduit: separate RQ_CB_CONDUIT [H])
    require 
      conduit.has_answer_for (Current)
    do 
      Result := conduit.remove (Current) 
    end
end
\end{lstlisting}
\caption{Sender endpoint class for message queue and callback connector
           \label{async_queue_callback}}

\end{figure}
\end{techreport}

\begin{techreport}
\figref{async_queue_callback} shows the
endpoint with which the sender will
transmit its message and receive
the callback. One wait condition indicates that there is space in the queue to send,
and the other indicates that there is a callback waiting for the sender.
The interface of this endpoint is simply the
\eif{send} and \eif{accept} procedures,
as seen in \figref{fig:scoop_queue_class_diagram}.
\end{techreport}

\begin{techreport}
\subsection{Integrating components and connectors}
\end{techreport}

Since connectors come in three parts: sender/receiver endpoints and the conduits,
any component that wants to use a connector
must have access to the connector's endpoint functionality. This can either be done by creating an endpoint object,
or inheriting from the appropriate endpoint class. Because the conduits are an implementation detail of the endpoints,
components do not need a direct reference to the conduits.
\begin{techreport}

An example where the conduit is given as an argument
to the component is given in \figref{fig:print_component}.
The \eif{PRINTER} class is responsible for accepting strings
from the connector and writing them to a physical piece of paper.
Upon creation the printer is given the conduit on which to receive the data, 
and at runtime it continually waits on the
endpoint's \eif{receive} method for new data.
The other end of the communication is a
processor that is generating the print jobs.
In this case it also initially sets up the printer,
although this is not a requirement, a third party could create both
\eif{COMPUTER} and \eif{PRINTER}.
The setup can be seen in \figref{fig:computer_send}.

\begin{figure}[h]
\begin{lstlisting}[language=SCOOP]
class PRINTER
create make
feature {NONE}
  receiver: SB_RECEIVER [separate STRING]
  make (conduit: separate SB_CONDUIT [separate STRING])
    do 
      create receiver.make (conduit) 
    end
  print (str: separate STRING) do ... end
feature
  start
    do 
      print (receiver.receive) 
    end
end
\end{lstlisting}
  \caption{Printer component}
  \label{fig:print_component}
\end{figure}

\begin{figure}[h]
  \begin{lstlisting}[language=SCOOP]
print_send_ep: SB_SENDER [separate STRING]
setup
  local
    print: separate PRINTER
    conduit: separate SB_CONDUIT [separate STRING]
  do
    create conduit
    create print_send_ep.make (conduit)
    create print.make (conduit)
  end
process
  do 
    print_send_ep.send (sep_string ("...")) 
  end
  \end{lstlisting}
  \caption{\eif{COMPUTER} sending data to \eif{PRINTER}}
  \label{fig:computer_send}
\end{figure}
\end{techreport}

\subsection {Mapping to other concurrent languages}
\label{sec:mapping-other}

In general, the COMET design patterns should be implementable by a
variety of concurrent object-oriented languages besides SCOOP.  The
choice of language depends on the requirements of the final system
(e.g., execution guarantees).

As an example of another mapping, we consider an alternative
implementation for Java.  While similar, the two implementations
differ in several ways.  Firstly, the Java system must rely on
programmer competence to ensure data race freedom; this is a manual and
generally difficult task.  As all data is potentially shared in Java
programs, one must determine which data is read/written by more than
one thread, and protect it accordingly.  SCOOP does not require such
manual tasks, as it ensures data race freedom by construction. In SCOOP it
is impossible to access the data from both the sending and receiving
side, while in Java this is possible.

Secondly, the concurrent data structures used 
(such as the concurrent queues used in the connectors)
require different implementation styles due to the language.
Although Java contains first-class support for concurrent programming
through the inclusion of monitors for every object,
implementing a concurrent queue still requires a measure of caution.
For example, implementing a condition variable requires the
identification of the condition,
waiting on the object that embodies the condition,
and having a corresponding signal in another routine that
is fired only when the condition is true.
All of these actions have the possibility to introduce an error.
In SCOOP, these manual tasks 
(condition identification, wait, and notification) 
are rolled into a single concept: the wait condition.
Signaling and waiting happen automatically, reducing the problem
to only ensuring that the correct precondition is used.

Java and SCOOP (in its current implementation) differ in expressivity
in one important case: SCOOP can successfully implement all but one of
COMET's suggested behavioral stereotypes (no ``multi-read entities'');
Java on the other hand has no problems with multiple readers. 
This is not a special characteristic of the SCOOP model:
this limitation is shared by all actor and message-passing concurrency models,
where shared memory is not considered.
Work is in progress to remove this limitation in SCOOP by
allowing multiple pure functions to access an object concurrently; 
purity of the function is determined by
either programmer annotations or static analysis.


\section{Case study}
\label{sec:case-study}
\begin{figure*}[htb]
  \centering
  \begin{tabular}{@{}c}
\subfloat[Design of the ATM system. Only one ATM with one customer is shown in detail; however, the server can be connected to multiple ATMs. To save space, the arrows omit the direction of the communication as done in \figref{fig:connectors for communication patterns}; instead, the names contain this information.]{
  \includegraphics[width=\textwidth]{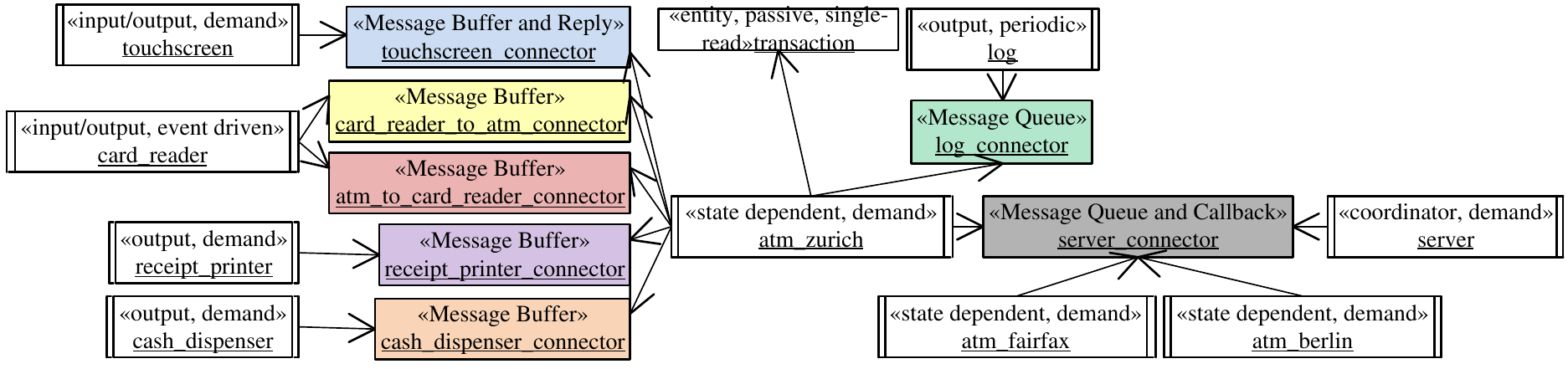}
  \label{fig:ATM system design}
}\\
\subfloat[SCOOP implementation of the ATM system. The boxes group objects handled by the same processor. Endpoint objects have the suffix {\lstinline[language=SCOOP]!ep!}. The names of the endpoint and conduit objects indicate the direction of the communication; for example, the atm\_receive\_ep object queries the touchscreen\_send\_conduit object to receive a message from the ATM. The colors link the connectors in \figref{fig:ATM system design} to the resulting connector objects in this figure.
]{
  \includegraphics[width=\textwidth]{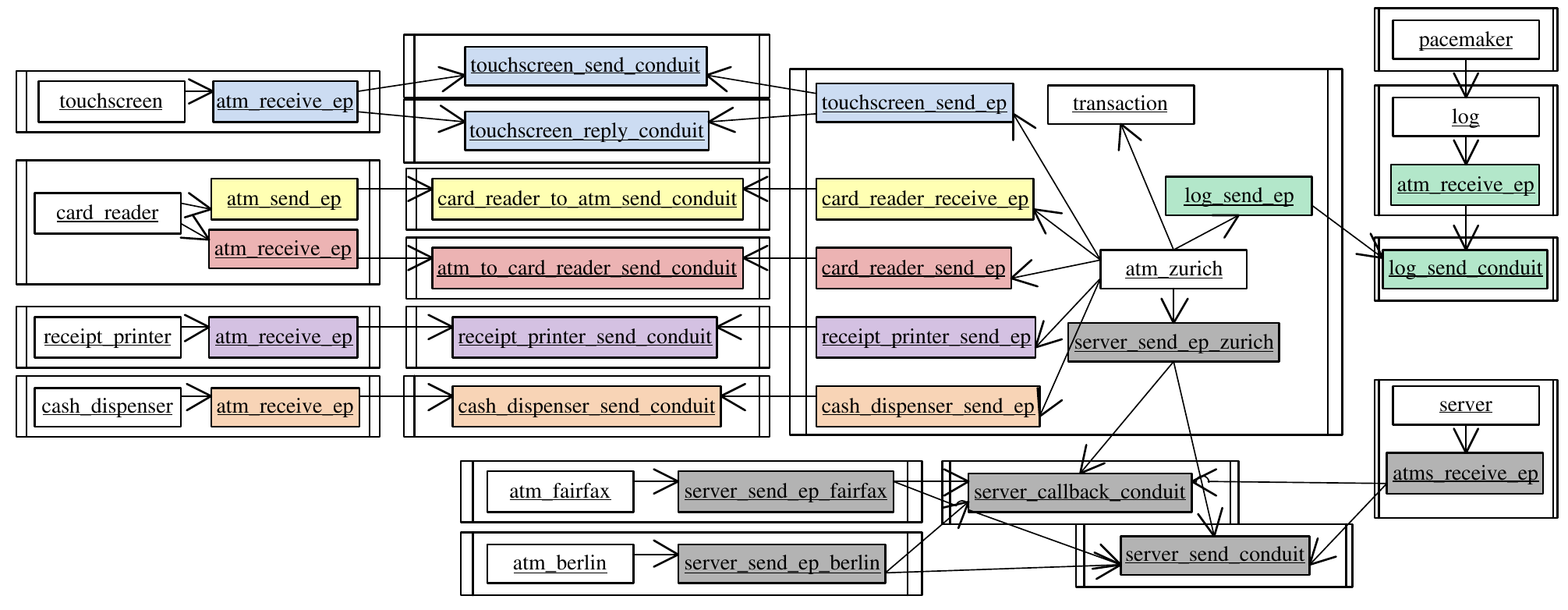}
  \label{fig:ATM system implementation}
}    
  \end{tabular}
  \caption{Design and implementation of the ATM system}
\end{figure*}

This section applies the suggested development method to an ATM system, \cite{gomaa:2000:designing}, 
shown in \figref{fig:ATM system design}. 
\begin{techreport}
The full implementation of the case study, together with the design pattern implementations, can be found online at~\cite{design_patterns_github:2012:ETH}.
\end{techreport}
\begin{conference}
The full implementation of the case study, together with the design pattern implementations, can be found online at~\cite{design_patterns_github:2012:ETH}.
\end{conference}

We chose the ATM system as an example, as it employs a wide spectrum of communication patterns: the synchronous patterns for an ATM's I/O devices and the asynchronous patterns for logging and the communication with the server. In particular, the example covers all of the connectors defined by COMET.

\begin{techreport}
In the ATM system, a central server manages the pins and numbers of the ATM cards, the mapping of ATM cards to bank accounts, and the bank accounts themselves. The server communicates with a number of ATMs over a message queue and callback connector. It waits for a request from an ATM and then processes the request: it either validates a pin, withdraws money, transfers money, or returns the balance of a bank account. Each ATM is state dependent and coordinates a number of simple components: its touchscreen, its card reader, its receipt printer, and its cash dispenser; a passive transaction keeps track of the customer's data. The ATM and the card reader are connected to each other over two message buffers: one for the initial \emph{card inserted message} from the card reader to the ATM and one for the \emph{return card} message from the ATM to the card reader. Each ATM additionally keeps a log to record important events; the log periodically reads from a message queue and saves the events to a permanent storage.

If the ATM design should accommodate more than one server,
then the server connector can simply be used to dispatch
the requests to multiple severs, rather than just one.
In this way, the system can achieve greater scalability due to the concurrent processing of requests.
\end{techreport}

\subsection{Applying the design pattern implementations}
The design pattern implementations from \secref{sec:implementation} can be used to build a SCOOP implementation of the ATM system. Each active component in the design becomes a component object handled by a separate processor; passive components become component objects handled by one of the processors for an active component object. The class of a component object inherits from the framework class that corresponds to the component's stereotype. For instance, the active ATM component becomes an object handled by a separate processor, and the passive transaction component becomes an object handled by the same processor as the ATM object. The ATM is a state dependent controller, and hence the class of the ATM object inherits from the corresponding framework class. \figref{fig:ATM system implementation} illustrates the result of this mapping.

\begin{techreport}
Each connector becomes either one or two conduit objects on separate processors, as described in \secref{sec:implementation}. For instance, a message queue and callback connector becomes a send conduit and a reply conduit on two separate processors. Whenever two components are connected over a connector, the resulting component objects each instantiate a non-separate endpoint object whose class inherits from the corresponding endpoint class. For instance, all ATMs share a message queue and callback connector to communicate with the server. Each ATM object instantiates an endpoint object of type \lstinline[language=SCOOP]!RQ_SENDER! to implement the sending endpoint of the connector; the server instantiates an endpoint object of type \lstinline[language=SCOOP]!RQ_RECEIVER! to implement the receiving endpoint.
\end{techreport}

\subsection{Implementing interconnections}
The design in \figref{fig:ATM system design} describes the interconnection of the components and connectors; however, it leaves the realization of these interconnections to the implementation. The root object is suitable to setup the objects representing control components, i.e., the server object and the ATM objects. To do so, the root object first creates the conduit objects that connect these control component objects. It then creates the control component objects and passes the conduit objects during creation; the component objects can then create local endpoint objects. 
\begin{conference}
After creation, the root object starts the new component objects.
\end{conference}
\begin{techreport}
After creation, the root object starts the new component objects, as shown in \figref{fig:atm system root class}. To be brief, type parameters for the connectors are omitted, and variable names are shortened.

\begin{figure}[htb]
\begin{lstlisting}[language=SCOOP]
class APPLICATION create make feature	
	make
		local
			server: separate SERVER
			atm_f, atm_z, atm_b: separate ATM
			s_send_conduit: separate RQ_SEND_CONDUIT
			s_cb_conduit: separate RQ_CB_CONDUIT
		do
			-- Create the connector objects.
			create s_send_conduit.make
			create s_cb_conduit.make
			-- Create the component objects.
			create server.make (s_send_conduit, s_cb_conduit)
			create atm_f.make (s_send_conduit, s_cb_conduit)
			create atm_z.make (s_send_conduit, s_cb_conduit)
			create atm_b.make (s_send_conduit, s_cb_conduit)
			-- Start the component objects.
			start_server (server)
			start_atms (atm_f, atm_z, atm_b)
		end
end
\end{lstlisting}
  \caption{Root class of ATM system}
  \label{fig:atm system root class}
\end{figure}
\end{techreport}

Each object representing a controlled component can be created by the controlling object. To do so, the control object first creates the conduit objects for the connectors along with local endpoint objects. It then creates the controlled object using the conduit objects. For instance, each ATM object creates a touchscreen object, a card reader object, a receipt printer object, a cash dispenser object, and a log object on separate processors.
\begin{techreport}
For each of these, the ATM object uses an endpoint object with suffix \lstinline[language=SCOOP]!ep!, as shown in \figref{fig:ATM class}.

\begin{figure}[htb]
\begin{lstlisting}[language=SCOOP]
class ATM inherit STATE_DEPENDENT create make
feature
	s_send_ep: RQ_SENDER
	ts_send_ep: RB_SENDER
	cr_receive_ep: SB_RECEIVER ...

	make (
		s_send_conduit: separate RQ_SEND_CONDUIT;
		s_cb_conduit: separate RQ_CB_CONDUIT
	)
			-- Create the ATM.
		local
			ts_send_conduit: separate RB_SEND_CONDUIT
			ts_reply_conduit: separate RB_REPLY_CONDUIT
			ts: separate TOUCHSCREEN ...
		do
			-- Create the connector objects. 
			create s_send_ep.make(s_send_conduit, s_cb_conduit)
			create ts_send_conduit.make
			create ts_reply_conduit.make
			create ts_send_ep.make (ts_send_conduit, ts_reply_conduit) ...		
			-- Create the component objects.
			create ts.make (ts_send_conduit, ts_reply_conduit) ...
		end
end
\end{lstlisting}
  \caption{Class for ATM}
  \label{fig:ATM class}
\end{figure}
\end{techreport}

\begin{techreport}
\begin{figure}[htb]
\begin{lstlisting}[language=SCOOP]
start
	local
		msg: MESSAGE
		t: TRANSACTION
	do
		from until stop loop
			-- Wait for message from card reader object.
			msg := cr_receive_ep.receive
			create t.make; t.set_card_number (msg.card_number)
			-- Get pin from customer.
			msg := ts_send_ep.send (create {MESSAGE}.get_pin)
			t.set_pin (msg.pin)
			-- Validate pin on server.
			s_send_ep.send (create {MESSAGE}.validate_pin (t.card_number, t.pin))
			msg := s_send_ep.accept
			if msg.has_succeeded then
				-- Get customer request.
				msg := ts_send_ep.send (create {MESSAGE}.get_rqst)
				if msg.is_withdrawal then ...
				else if msg.is_transfer then ...
				else if msg.is_balance_retrieval then ... end
			else ... end
		end
	end
\end{lstlisting}
  \caption{Starting feature of ATM}
  \label{fig:ATM start feature}
\end{figure}
\end{techreport}

\subsection{Implementing interactions}
The interactions between components can be implemented in the \lstinline[language=SCOOP]!start! features of the component objects. For instance, an ATM object executes a loop where each iteration begins with a message from the card reader object. Upon receiving this message, the ATM object retrieves the pin from the touchscreen object, validates the pin with the server object, logs the result, and then proceeds according to the customer's choice. The ATM object stores the intermediate results into a transaction object. 
\begin{conference}
For this purpose, the ATM object creates the transaction object on its own processor when it receives a message.
\end{conference}
\begin{techreport}
For this purpose, the ATM object creates the transaction object on its own processor when it receives a message, as shown in \figref{fig:ATM start feature}.
\end{techreport}
 
The server object executes a similar loop: it waits for messages from one of the ATM objects and acts according to the message's nature. It then reports whether or not it was able to process the request successfully.

\subsection{Discussion}
The case study was a manual effort; the proposed development method has however the potential for automation. Deriving an implementation from the UML design involves the following steps:

\begin{enumerate}
\item Generate one class for each component. The class inherits from the framework class corresponding to the component's stereotype. For each of the component's connectors, the class has an attribute for the connector's endpoint object; for each passive component, the class has an attribute as well. The class has a creation procedure to initialize these attributes. For each connector, the creation procedure takes the connector's conduit objects as arguments and uses them to initialize the endpoint object. Finally, the creation procedure creates a non-separate component object for each passive component.
\item Generate one root class. The root class first creates the conduit objects for each connector. It then creates a component object on a separate processor for each active component. It links the component objects according to the design by passing the conduit objects during construction. Lastly, the root class triggers the execution of all created component objects.
\item In each component class, add code that implements the component's specification. This code contains the interactions between the component objects over the connector objects.
\end{enumerate}

The first and second step can be automated because the necessary information is available in the design. However, the design does not capture the application logic. Hence, a tool can only generate templates to which developers must manually add the logic in the third step.

The systematic mapping of components and connectors to objects and processors ensures traceability. Each object in the application can be mapped to exactly one component or connector; vice versa, each component and connector can be mapped to a distinct set of objects.

\section{Related work}
\label{sec:related-research}

Software design patterns provide a tried and tested solution to a
design problem in the form of a reusable template, which can be
used in the design of new software applications. Software
architectural patterns~\cite{buschmann:1996:pattern} 
address the high-level design of the software architecture
\cite{shaw:1996:software,taylor:2009:software},
usually in terms of components and connectors. These include 
widely used architectures~\cite{bass:2003:software} such as client/\-server
and layered architectures. 
Design patterns
\cite{gamma:1995:design} address smaller reusable designs than
architectural patterns 
in terms of communicating objects and
classes customized to solve a general design problem in a particular
context. The patterns described in this paper are aimed at developing concurrent applications and are hence different from patterns for sequential applications.

Component technologies~\cite{taylor:2009:software, szyperski:2002:component} have been developed
for distributed applications. Examples of this technology include client-side Java Beans and server-side Enterprise Java Beans (EJB). A bean is a reusable component that typically
consists of multiple objects. EJBs encapsulate application logic that
can be accessed by client beans. EJB containers provide system-wide
services such as message communication and transaction management.

Patterns for concurrent and networked objects are described in \cite{schmidt-stal-rohnert-buschmann:2000:patterns_for_concurrency}; the patterns are comprehensive and largely oriented to middleware development.  However, these patterns are not used to systematically derive a concurrent program from a design, as it is the case in our approach.

Fliege et al.~\cite{fliege-et-al:2005:DSR:1122889.1122895} present design patterns to detect fail-silent components in concurrent real-time systems and use them to implement an airship control system. 
Bellebia and Douin~\cite{bellebia-douin:2006:applying} use design patterns to develop a middleware for embedded systems. Some of their patterns also address component structures and communication between concurrent components. In contrast to our work, the design patterns in these two works focus on failure detection and middlewares, respectively, and do not capture general interactions between concurrent components. 
Pettit and Gomaa~\cite{pettit:2006:modeling} represent UML models using colored Petri nets to conduct behavioral analyses (e.g., ~timing behavior). Our work focuses on obtaining an executable system with built-in behavioral guarantees; in future work, the two approaches could be combined to offer both safe execution and advanced behavioral analyses of the model. 

A number of approaches address the generation of code from design patterns, e.g.,~\cite{budinsky-et-al:1996:automatic,oshtsuki-et-al:1999:source}. These approaches reduce the work needed to apply existing design patterns to a program. However, they do not generate code for designs of entire concurrent programs.

Shousha et al.~\cite{shousha-et-al:2008:model} present an approach to detect data races in UML models of concurrent programs. Our approach prevents data races entirely because SCOOP programs are data race free by design. Hence, it is unnecessary to perform such an analysis in our approach.


\section{Conclusion}
\label{sec:conclusion}

With the increasing need of concurrency, offering adequate support to developers in designing and
writing concurrent applications has become an important challenge. The
approach taken in this paper is to base such support on widely used
architectural modeling principles, namely UML with the COMET method,
which should simplify adoption in industrial settings. We defined a
mapping of COMET's behavioral design patterns into SCOOP programs and
demonstrated with a case study that using this approach entire
concurrent UML designs can be systematically mapped to executable
programs. Choosing SCOOP rather than another concurrent language has
the important benefit that the resulting programs inherit SCOOP's
execution guarantees, i.e.,\ are data-race free by construction.


For future work, it would be interesting to integrate our method with
other approaches based on UML and the CO\-MET method, giving rise to a
more comprehensive framework with additional analyses of
concurrent designs, e.g.,\ concerning their timing
properties~\cite{pettit:2006:modeling}. In the long term, we would
like to provide an automated method to translate UML concurrent
software architecture designs to an implementation. We are also
planning to implement further patterns, 
for example event-based, transaction-based, and brokered patterns,
used exclusively in distributed communication.

\subsubsection*{Acknowledgments}
The research leading to these results has received funding from the European Research Council under the European Union's Seventh Framework Programme (FP7/2007-2013) / ERC Grant agreement no.\ 291389, the Hasler Foundation, and ETH (ETHIIRA). 




\begin{conference}
\bibliographystyle{abbrv}
\end{conference}
\bibliography{bibfile}

\end{document}